\documentclass[aps,prl,twocolumn,showpacs,preprintnumbers,amsmath,amssymb]{revtex4}
\usepackage{graphicx}% Include figure files
\usepackage{dcolumn}% Align table columns on decimal point
\usepackage{mathrsfs}
\usepackage{bm}% bold math
\usepackage{natbib}

\usepackage{amsfonts}
\usepackage{amssymb}
\usepackage{amsmath}
\usepackage{color}
\usepackage{float}

\begin{document}
\preprint{}

\title{Dressed-State Polarization Transfer between Bright \& Dark Spins in Diamond}

\author{C. Belthangady$^{1,2}$, N. Bar-Gill$^{1,2}$, L. M. Pham$^{3}$, K. Arai$^{4}$, D. Le Sage$^{1,2}$, P. Cappellaro$^{5}$, and R. L. Walsworth$^{1,2}$}

\email{rwalsworth@cfa.harvard.edu}
\address{$^{1}$
Harvard-Smithsonian Center for Astrophysics, 60 Garden Street Cambridge MA-02138, USA.}
\address{$^{2}$Department of Physics, Harvard University, Cambridge, 
Massachusetts 02138, USA.}
\address{$^{3}$
School of Engineering and Applied Sciences, Harvard University, Cambridge, Massachusetts 02138, USA.}
\address{$^{4}$
Department of Physics, Massachusetts Institute of Technology, Cambridge, Massachusetts 02139, USA. }
\address{$^{5}$Nuclear Science and Engineering Department, Massachusetts Institute of Technology, Cambridge, Massachusetts 02139, USA. }
\date{\today}

\begin{abstract}
Under ambient conditions, spin impurities in solid-state systems are found in thermally-mixed states and are optically ``dark", i.e., the spin states cannot be optically controlled. Nitrogen-vacancy (NV) centers in diamond are an exception in that the electronic spin states are``bright", i.e., they can be polarized by optical pumping, coherently manipulated with spin-resonance techniques, and  read out optically, all at room temperature. Here we demonstrate a dressed-state, double-resonance scheme to transfer polarization from bright NV electronic spins to dark substitutional-Nitrogen (P1) electronic spins in diamond.  This polarization-transfer mechanism could be used to cool a mesoscopic bath of dark spins to near-zero temperature, thus providing a resource for quantum information and sensing, and aiding studies of quantum effects in many-body spin systems.
\end{abstract}


\maketitle

Nitrogen-vacancy (NV) color centers in diamond have attracted wide interest recently for applications in quantum information \cite{Quantum_Comp_Rev} and magnetometry \cite{Taylor,Degen,Maze,Gopi}. Key characteristics of NV centers are their long electronic spin coherence lifetimes ($\sim$ ms) and their optically ÒbrightÓ nature: i.e., the ability to  prepare and read out NV spin states optically at room temperature. The diamond lattice is also host to many other dark spin impurities \cite{Loubser} i.e., electronic and nuclear spins that cannot be initialized or read out optically, and under ambient conditions are found in a thermal mixture of spin states. Fluctuating magnetic fields associated with this thermal bath of dark spins are a major source of decoherence for NV spins \cite{deLange, Colm, Boris, Nir, Hanson-spin-bath, Linh}. A key challenge is to transfer polarization controllably from bright NV spins to dark spins both within and outside the diamond lattice, which can mitigate NV spin decoherence, convert the dark spins into a resource for quantum information \cite{Bose,Norman_Yao} and sensing \cite{Goldstein, Cai_Fedor} and aid the study of quantum fluctuations and dynamics in many-body spin systems. Here, we use a dressed-state, double-resonance scheme to realize such polarization transfer from bright NV electronic spins to more abundant, thermally-mixed, dark substitutional-Nitrogen (P1) electronic spins in room temperature diamond. Such optically-controlled polarization transfer could be used to effectively cool a wide range of dark spin ensembles at room (and arbitrary) temperature.

\begin{figure}[h]
\begin{center}
\includegraphics*[width=3.5 truein]{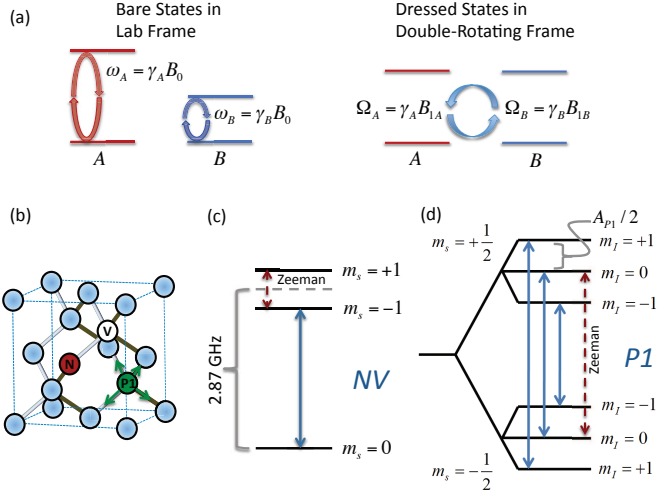}
\end{center}
\caption{(color online.) Schematic of dressed-state polarization transfer and energy-level diagrams. (a) Two spins with dissimilar Zeeman splittings cannot exchange energy in the lab frame. When driven resonantly (red and blue circular arrows), energy exchange becomes possible. Dressed states in the double-rotating frame are separated by the spin Rabi frequencies controlled by the respective driving fields, $B_{1A,B}$. When the Rabi frequencies are matched, i.e., $\Omega_{A}=\Omega_{B}$, energy conserving spin flip-flops can occur. (b) Nitrogen-vacancy (NV) and substitutional-Nitrogen (P1) defects in diamond. Green arrows show the four possible orientations of the P1 hyperfine axis. (c) Ground electronic spin states of the negatively charged NV center. Degeneracy of the $|m_{s}=\pm1\rangle$ states is lifted by application of a static magnetic field, $B_{0}$, aligned parallel to the NV axis. (d) Ground electronic spin states of the P1 defect. For $B_{0}$ oriented along the $\langle111\rangle$ crystal direction, the hyperfine splitting is $A_{P1}=114~\rm{MHz}$ when the hyperfine axis is parallel to $B_{0}$, and  $A_{P1}= 90~\rm{MHz}$ for the three other orientations.}
\label{fig:HHScheme}
\end{figure}

Polarization of a solid-state dark-spin bath is possible at cryogenic temperatures and magnetic fields of several Tesla \cite{Takahashi}, but has not been realized at room temperature. In previous experiments, electronic spin polarization transfer between a single NV and a single proximal P1 center was demonstrated by tuning the static magnetic field, $B_{0}$ to a level anti-crossing \cite{Gaebel,Hanson_pol}. However, when a diverse set of dark spins with different gyromagnetic ratios and/or hyperfine couplings are present, as is the case in typical diamond samples, such tuned $B_{0}$ polarization transfer fails. Using dressed-states for spin polarization transfer is robust in that it works at room temperature for any number of dark spin species, with magnetic fields that are both low ($\sim$mT) and user-specified, and provides fast optical control. The P1 center, with its strong and orientation-dependent hyperfine coupling, exhibits a set of five optically-dark electron spin resonance (ESR) transitions, and thereby serves as an ideal platform to demonstrate dressed-state polarization transfer from bright NV spins. As described below, we measured the spectral width of the NV/P1 double-resonance interaction and the rate of polarization transfer between bright and dark spins, and thus established the conditions for optically-controlled polarization (i.e., effective cooling) of mesoscopic dark spin ensembles in room-temperature diamond.

The essence of dressed-state polarization transfer is illustrated in Fig.\ref{fig:HHScheme}(a). Two types of spins, A and B, with different gyromagnetic ratios ($\gamma_{A},\gamma_{B}$) have unequal Zeeman splittings when placed in an external magnetic field ($B_{0}$) and therefore cannot exchange energy in the lab frame. However, if both sets of spins are driven resonantly at their respective Larmor frequencies with transverse electromagnetic fields such that their $\it{Rabi}$ frequencies are equal (analogous to the Hartmann-Hahn matching condition \cite{Hartmann-Hahn}), then in a double-rotating frame the dressed states of the two spins have equal energy separation and energy transfer becomes possible. If one of the species is spin-polarized, this polarization may be transferred to the other species by means of a resonant flip-flop process in the double-rotating frame mediated by their mutual magnetic coupling. Dressed-state polarization transfer has been studied for nuclear spins in bulk as well as nanoscale ensembles \cite{Rugar}. Modified dressed-state schemes have also been used to transfer thermal polarization from electronic to nuclear spins in bulk ensembles \cite{Reynhardt}. Here, we apply this technique to dissimilar electronic spins and demonstrate polarization transfer from optically polarized spins of lower abundance (NV centers) to dark (P1) spins of higher abundance.

The NV center (Fig.\ref{fig:HHScheme}(b)) consists of a substitutional Nitrogen atom and an adjacent vacancy in the diamond lattice. The ground state of the negatively charged NV center (the focus of the present work), shown in Fig.\ref{fig:HHScheme}(c), is an electronic spin triplet (S=1) with a zero-field splitting of 2.87 GHz. The degeneracy of the $|m_{s}= \pm 1\rangle$ states may be lifted with an external $B_{0}$ field and spin transitions between the Zeeman states can be driven by means of microwave radiation. Optical excitation at wavelengths shorter than the zero-phonon line at 638nm leads to polarization of the NV electronic spin into the $|m_{s}=0\rangle$ state. The P1 center consists of a Nitrogen atom that has replaced one of the carbon atoms in the lattice (Fig.\ref{fig:HHScheme}(b)). A Jahn-Teller distortion of one of the N-C bonds leads to an anisotropy in the hyperfine interaction between the P1 electronic spin and its nuclear spin (predominantly $^{14}N$ with $I=1$), Fig.\ref{fig:HHScheme}(d) gives the energy level diagram of the P1 center and the three allowed electronic spin transitions. The hyperfine energy splitting is determined by the angle between the anisotropic hyperfine axis (the distorted N-C bond direction) and the static magnetic field, so that when $B_{0} \parallel \langle111\rangle$ crystal direction, the ESR spectrum of the P1 center comprises a set of five lines \cite{Smith, Takahashi}. 

If the dark, thermally-mixed P1 spins are brought into resonance with spin-polarized NVs, then polarization can be transferred from NVs to P1s mediated by the magnetic dipolar coupling with Hamiltonian:
\begin{equation}
\label{hdip}
H_{dip}=\mathcal{D}_{NV,P1}[S^{NV}_{z}S^{P1}_{z} -\frac{1}{4\surd{2}}(S^{NV}_{+}S^{P1}_{-}+S^{NV}_{-}S^{P1}_{+})].
\end{equation} 
Here ${\bf S}^{NV}$ and ${\bf S}^{P1}$ are the spin operators for the NV and P1 spins respectively, $\mathcal{D}_{NV,P1}=\mu_{0}\gamma_{NV}\gamma_{P1}\hbar^{2}(1-3\cos^{2}\theta)/4\pi r^{3}$ where $r$ and $\theta$ are co-ordinates of the position vector connecting the spins, and $\gamma_{NV} \approx \gamma_{P1} = 2\pi \times$ 2.8 MHz/gauss. One method to bring the P1 and NV spins into resonance is to tune the static magnetic field to $B_{0}= \rm{512~gauss}$. At this value of the field, the energy difference between the $|m_{s}=0\rangle$ and $|m_{s}=-1\rangle$ NV spin states is equal to that between the  $|m_{s}=-1/2, m_{I}=0\rangle$ and $|m_{s}=1/2, m_{I}=0\rangle$ P1 spin states, and thus the second term of $H_{dip}$ induces a resonant flip-flop process. However, this tuned $B_{0}$ technique fails when several P1 spins are strongly coupled to each NV (there are $\sim$ 10 to 1000 P1s per NV in typical diamond samples),  because at a given $B_{0}$ field only one of the five P1 hyperfine transitions is on resonance with the NV. For the more generally-applicable dressed-state scheme introduced here, the NV and P1 spins are locked in a  direction transverse to the static $B_{0}$ field by controllable, continuous driving at microwave and RF frequencies, respectively. When the drive Rabi frequencies are matched, the NV and P1  spins are brought into resonance in a double-rotating frame, such that the dipolar Hamiltonian consists of only the first term of equation (\ref{hdip}) \cite{Slichter}. This term $\sim S^{NV}_{z}S^{P1}_{z}$ is perpendicular to the spin-locking direction and can therefore induce flip-flops between the P1 and NV spins. The primary advantage of dressed-state polarization transfer is  that by driving each P1 hyperfine transition on resonance  with a multi-frequency RF field all five P1 transitions can be simultaneously brought into resonance with the NV in the rotating frame, allowing polarization transfer from the NV to all P1 spins. By direct analogy, polarization can be transferred from bright NV spins to other dark  (nuclear or electronic) spins with different Zeeman energies by driving each spin transition on resonance, simultaneously, and with matched Rabi frequencies. Other advantages of dressed-state polarization transfer include: (i) the ability to switch on and off the polarization mechanism on nanosecond time scales (by rapid switching of the RF and/or microwave fields), and (ii) the ability to transfer polarization at any value of $B_{0}$.

\begin{figure}[t]
\begin{center}
\includegraphics*[width=3.5 truein]{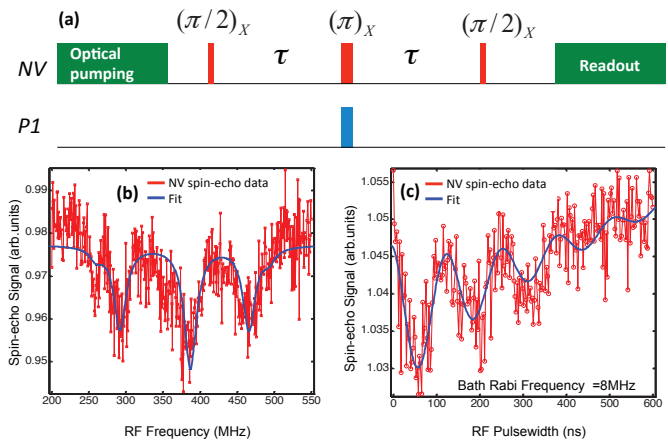}
\end{center}
\caption{(color online.) Measured P1-bath ESR spectrum and Rabi oscillations. (a) NV optical and microwave (red) spin-echo sequence with a fixed free precession time of 350 ns is used to measure the P1 ESR spectrum and Rabi oscillations, for $B_{0}= 128 ~\rm{gauss}$. Coincident with the NV $\pi$ pulse a 65 ns-long RF pulse (blue) is applied to the P1 bath spins. (b) P1-bath ESR spectrum is given by measured NV spin-echo fluorescence signal and NV/P1 dipolar coupling as the width of the P1 RF pulse is kept fixed and its frequency is swept. The $|m_{s}=-1/2, m_{I}=0\rangle \leftrightarrow |m_{s}=1/2, m_{I}=0\rangle$ transition for all four P1 orientations yields the observed central NV fluorescence dip. The satellite dips are from the $|m_{s}=-1/2, m_{I}=\pm 1\rangle \leftrightarrow |m_{s}=1/2, m_{I}=\pm 1\rangle$ transitions of off and on-axis P1s. The small dip appearing on the shoulder of the central dip arises from other dark spins in the lattice \cite{Loubser, Meriles}. Solid line represents a fit to a set of five Gaussians. (c) With the frequency of the RF signal set to the central P1 resonance dip, the width of the RF pulse is swept to yield P1-bath Rabi oscillations via effect of NV/P1 dipolar coupling on NV spin-echo fluorescence measurements. Solid line is a fit to a decaying sinusoid.}
\label{fig:ESR_Rabi}
\end{figure}
\begin{figure}[t]
\begin{center}
\includegraphics*[width=3.5 truein]{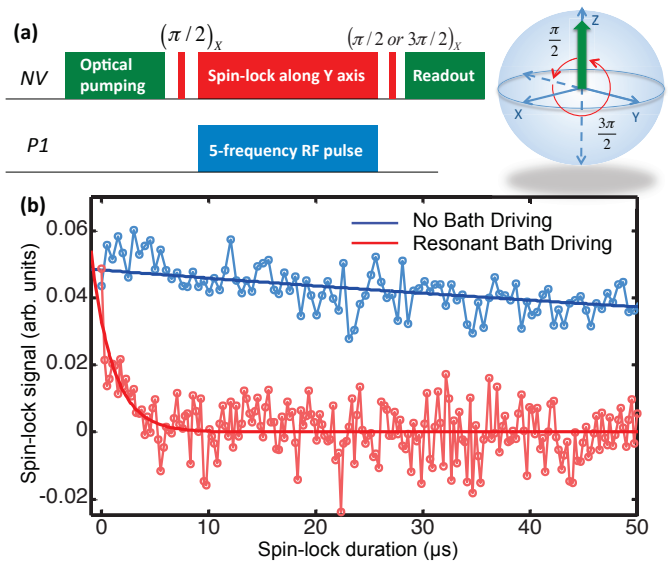}
\end{center}
\caption{(color online.) Measured dressed-state NV/P1 polarization transfer. (a) NV optical and microwave spin-lock sequence, and timing of 5-frequency RF pulse to drive P1 spin bath. After initializing NVs to the the $|m_{s}=0\rangle$ state (Z-axis) with a $2\mu s$ optical pumping pulse, $\pi/2$ and $3\pi/2$ microwave pulses are applied along the $X$ axis to rotate the spin to/from the equator of the Bloch sphere, and a microwave spin-locking pulse of variable duration is applied along the $Y$ axis. To bring the P1 bath into dressed-state resonance with the NV spins, a 5-frequency RF pulse is applied during the spin-lock duration. (b) Measured decay of NV spin-lock signal as a function of spin-lock duration. When the P1 bath is not driven we record the blue trace. Driving the P1 bath simultaneously at all 5 ESR resonances such that the collective P1 bath Rabi frequency equals the NV Rabi frequency (8 MHz) gives the red trace, indicating strong NV/P1 polarization transfer. Solid lines represent fits to decaying exponentials.}
\label{fig:HHDecay}
\end{figure}

Our experiments were performed using a home-built confocal microscope to interrogate a small ensemble ($\sim 10^{4}$) of NVs in an NV-dense diamond sample (see Supplementary Material). Applying Ramsey and Hahn-echo pulse sequences \cite{Slichter} to the NV $|m_{s}=0\rangle  \leftrightarrow  |m_{s}=-1\rangle$ transition, we determined $T^{*}_{2}= 110(10)~\rm{ns}$ and $T_{2}=1.6(1) ~\mu s$. To identify the resonance frequencies of the P1 spin-bath we used the double-electron-electron-resonance (DEER) sequence shown in Fig.\ref{fig:ESR_Rabi}(a) \cite{Hanson-spin-bath}. The frequency of the applied RF signal was swept to record the P1 bath-ESR spectrum via the effect on the NV spin-echo fluorescence signal, as shown in Fig.\ref{fig:ESR_Rabi}(b). In addition to the P1 spectrum we found a DEER signal from an unknown electronic spin impurity (which is ubiquitous in all diamond samples we measured). We next set the frequency of the RF pulse to that of the central fluorescence dip, varied the width of the pulse, and recorded the NV spin-echo signal as an indicator of P1-bath Rabi oscillations (Fig.\ref{fig:ESR_Rabi}(c)). We then used an RF source outputting five frequencies to drive each of the five P1 ESR transitions, and adjusted the power at each frequency such that the Rabi frequency of every transition was equal to 8 MHz. We were thus able to drive collective, synchronized Rabi oscillations of the entire P1 spin bath.

The pulse sequence we employed for dressed-state polarization transfer is shown in Fig.\ref{fig:HHDecay}(a). The
microwave power was adjusted to give an NV Rabi frequency of 8 MHz (matched to the P1 Rabi frequency). Measuring the NV spin-lock signal as a function of the duration of the spin-lock pulse with no RF signal applied, we found the rotating-frame spin-lattice relaxation time $T_{1,\rho}= 290(50) ~\mu s$. The blue trace of Fig.\ref{fig:HHDecay}(b) shows the first 50 $\mu s$ of this measured NV spin-lock signal. When the P1 spin bath was driven  such that the collective Rabi frequency was also 8 MHz, we observed a $\sim$ 100x faster decay of the NV spin-lock signal with $T_{1,\rho}= 2.0 (4)~\mu s$, as shown by the red trace of Fig.\ref{fig:HHDecay}(b). This dramatic effect demonstrates that in the rotating frame the NV spins are much more strongly coupled to the P1 spin bath under the condition of matched P1 and NV Rabi frequencies. Note that since the P1 spin bath is selectively recoupled to the spin-locked NVs,  $T_{1,\rho}$ approximately equals $T_{2}$ in the lab (non-rotating) frame.

\begin{figure}[t]
\begin{center}
\includegraphics*[width=3.5 truein]{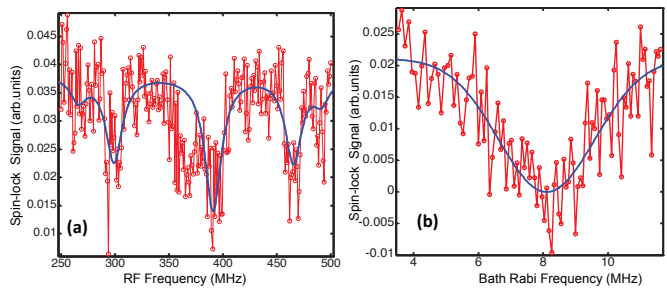}
\end{center}
\caption{(color online.) Measured spectrum of NV-spin polarization transfer to P1 bath spins. Here $B_{0}=$132 gauss and the NV Rabi frequency = 8 MHz. (a) A single-frequency RF signal, with power corresponding to P1 Rabi frequency  of 8 MHz, is swept across the P1 bath resonances. When the RF frequency equals that of a P1 spin bath resonance, the NV spin-lock signal is reduced. Solid line is a fit to a set of five Gaussians. (b) NV/P1 double resonance spectrum is measured by monitoring the NV spin-lock signal,  applying RF at the central P1 bath resonance dip, and sweeping the power of the RF signal. Solid line represents a Gaussian fit. The resonance is centered at 8.1(1) MHz and has a FWHM of 3.5(4) MHz.}
\label{fig:HHCurve}
\end{figure}

To verify that the rapid decay of the NV spin-lock signal was caused by transfer of polarization to the P1 spins, we set the NV Rabi frequency to 8 MHz and the spin-lock duration to 50 $\mu s$; drove the P1 spin bath at a Rabi frequency of 8 MHz (at a single RF frequency instead of all five); and slowly swept the RF frequency past the five P1-bath ESR resonances. Each time the RF frequency equaled that of a P1-bath resonance, we observed a drop in the NV spin-lock signal (Fig. \ref{fig:HHCurve}(a)) consistent with enhanced NV-spin polarization transfer to the P1-spin bath. Through this measurement we could also infer polarization transfer to the unknown electronic spin impurity. We next characterized the NV/P1 double resonance spectrum by setting the NV Rabi frequency to 8 MHz and the spin-lock duration to 50 $\mu s$; setting the RF signal frequency to that of the central ESR dip of the P1 spin bath; and scanning the RF power to tune the P1-bath Rabi frequency through the double resonance condition. When the RF power was such that the P1 and NV Rabi frequencies were equal (8 MHz), we observed a reduction in the NV spin-lock signal as shown in Fig.\ref{fig:HHCurve}(b). This NV/P1 polarization transfer resonance is broadened by dipolar coupling between P1 spins with width approximately equal to $1/\pi T^{*}_{2}$.

To apply this dressed-state polarization transfer scheme to polarize multiple P1 spins as a resource for quantum information, sensing, and metrology, repeated NV-initialization and polarization transfer cycles will be required. Also, the achievable P1 bath spin polarization will be limited by the P1 spin-lattice relaxation time, which is $\sim 1\rm{ms}$ in room temperature diamond \cite{Takahashi}. Since polarization transfer takes place within $\sim \rm{2} \mu s$,  as shown above, it should be possible to perform several hundred polarization cycles before the P1 spins relax.  Another important factor that will limit the achievable P1-bath spin polarization in steady-state is spin-diffusion between closely spaced P1 spins. This limitation may be overcome, for example, by restricting the size of the spin bath by  confining it within sub-micron structures \cite{Babinec}, or by realizing a higher ratio of optically bright  NV centers (polarization sources) to dark P1 spins to be polarized. 

In conclusion, we demonstrated that matching drive Rabi frequencies enables high-efficiency polarization transfer from optically bright NV electronic spins to nearby dark P1 electronic spins in room-temperature diamond, even when the strength of the NV/P1 coupling is not known or is inhomogenous. With optimized samples, such dressed-state polarization transfer will enable mesoscopic ensembles of dark spins to be polarized with optical control and at arbitrary magnetic fields, opening up several exciting possibilities.  Chains of polarized dark spins could be used to enable coherent state transfer between distant NVs \cite{Bose, Paola_state_transfer,Norman_Yao}. Environmentally-enhanced magnetometry, where the dark spins are used for field sensing, could increase the sensitivity of NV-based magnetometers \cite{Goldstein,Paola_EAM}. The polarization-transfer mechanism described here could also be extended to detect and polarize spins external to the diamond and thus enable single-molecule magnetic-resonance spectroscopy \cite{Rugar2004, Cai_Fedor} and diamond-based quantum simulations \cite{Cai_simulator}.

This work was supported by NIST, NSF and DARPA (QuEST and QuASAR programs). We gratefully acknowledge the provision of diamond samples by Apollo and helpful technical discussions with Stephen DeVience, Michael Grinolds, Patrick Maletinsky, Ashok Ajoy, Alexandre Cooper-Roy, Mikhail Lukin, Amir Yacoby, and Fedor Jelezko.\\

\end{document}